\documentclass[conference]{IEEEtran}
\IEEEoverridecommandlockouts
\usepackage{caption}
\usepackage{graphicx}
\usepackage{subfigure}
\usepackage{lipsum}
\usepackage{url}
\usepackage{verbatim}
\usepackage{todonotes}
\usepackage{color}

\usepackage{graphics}
\usepackage{amsmath,amsfonts, amssymb}
\usepackage{mathtools}
\usepackage{epstopdf}
\usepackage{amsthm}
\usepackage{bbm, bm}
\usepackage{algorithm}
\usepackage[noend]{algpseudocode}
\usepackage{dsfont}
\usepackage{cite}
\usepackage{multicol}
\usepackage{multirow}
\usepackage{array}
\usepackage{booktabs}
\usepackage[normalem]{ulem}
\useunder{\uline}{\ul}{}

\newcolumntype{P}[1]{>{\centering\arraybackslash}p{#1}}
\newcolumntype{M}[1]{>{\centering\arraybackslash}m{#1}}

\def\BibTeX{{\rm B\kern-.05em{\sc i\kern-.025em b}\kern-.08em
    T\kern-.1667em\lower.7ex\hbox{E}\kern-.125emX}}
\begin{document}

\title{Feedback is Good, Active Feedback is Better:\\ Block Attention Active Feedback Codes \\
\thanks{This work was partially supported by the European Research Council (ERC) through project BEACON (No. 677854).}
}

\author{\IEEEauthorblockN{Emre Ozfatura$^\dagger$, Yulin Shao$^\dagger$, Amin Ghazanfari$^\ddagger$, Alberto~Perotti$^\ddagger$, Branislav~Popovic$^\ddagger$, Deniz G{\"u}nd{\"u}z$^\dagger$}
\IEEEauthorblockA{$^\dagger$~Information Processing and Communications Lab (IPC-Lab), Imperial College London, London, UK \\ 
$^\ddagger$~Radio Transmission Technology Lab, Huawei Technologies Sweden AB, Kista 164-94, Sweden
}
}
\maketitle

\begin{abstract}

Deep neural network (DNN)-assisted channel coding designs, such as low-complexity neural decoders for existing codes, or end-to-end neural-network-based auto-encoder designs are gaining interest recently due to their improved performance and flexibility; particularly for communication scenarios in which high-performing structured code designs do not exist. Communication in the presence of feedback is one such communication scenario, and practical code design for feedback channels has remained an open challenge in coding theory for many decades. Recently, DNN-based designs have shown impressive results in exploiting feedback. In particular, \textit{generalized block attention feedback (GBAF) codes}, which utilizes the popular transformer architecture, achieved significant improvement in terms of the block error rate (BLER) performance. However, previous works have focused mainly on passive feedback, where the transmitter observes a noisy version of the signal at the receiver. In this work, we show that GBAF codes can also be used for channels with active feedback. We implement a pair of transformer architectures, at the transmitter and the receiver, which interact with each other sequentially, and achieve a new state-of-the-art BLER performance, especially in the low SNR regime.

\end{abstract}


\section{Introduction}
Designing error correction codes to combat channel noise has long been an intellectual pursuit of coding theorists and has important applications in wireless communications and data storage \cite{gallager1962low,berrou1993near,arikan2009channel,AlphaSeq}. Over the past few decades, many excellent coding schemes have been crafted by exploring their algebraic structures, such as the low-density parity-check (LDPC) code \cite{gallager1962low}, turbo codes \cite{berrou1993near}, polar codes \cite{arikan2009channel}, etc.

This paper investigates channel coding for a special class of channels: the feedback channel \cite{Shannon}. In addition to the conventional feedforward link from the transmitter to the receiver, the feedback channel provides an extra feedback link from the receiver to the transmitter, whereby the receiver can feedback information to the transmitter to aid feedforward data transmission. In 1956, Shannon \cite{Shannon} introduced the classical feedback channel model with unit-time delay, and proved that feedback from the receiver does not increase the channel capacity. In 1966, Schalkwijk and Kailath showed a surprising result that feedback does improve the reliability of feedforward transmission. The proposed feedback coding scheme, which is now known as the SK scheme \cite{SK1,SK2}, achieves super-exponential decay of error rate in the block length. Since then, designing good feedback codes, particularly amenable to practical implementation, has been an important open challenge in coding theory. 

Shannon, Schalkwijk and Kailath’s works consider only additive white Gaussian noise (AWGN) channels, and passive and noiseless feedback with unit-time delay. 
Later works focused on extending the feedback channel model to more general cases. When the feedback channel is noisy, Kim et al. \cite{Kim2007ISIT} proved that no linear code could achieve a positive communication rate with passive feedback. Ben-Yishai and Shayevitz \cite{ModuloSK} considered the active feedback scenario, in which the receiver can process its received signal before feeding it back to the transmitter, and proposed a modulo-SK scheme that achieves close-to-optimal performance when the feedback channel signal-to-noise ratio (SNR) is sufficiently larger than the feedforward channel SNR. In \cite{Gallager2} and \cite{Sahai}, the authors consider variable length coding, and in \cite{ozarow1984achievable,ozarow1984capacity,bross2008relay}, the feedback channel model is extended to multi-user channels, such as broadcast, multiple-access, and relay channels.

More recently, data-driven approaches, e.g., deep learning (DL), have been proposed for feedback code design \cite{deepcode,defc,drf,AttentionCode,gbaf}. In this approach, the channel encoder and decoder are parameterized by deep neural networks (DNNs), and the feedback communication system, including the feedforward and feedback channels, is modeled as part of an auto-encoder architecture.  By sampling channel realizations, one can jointly train the channel encoder and decoder in an end-to-end fashion, such that a bitstream fed into the communication system can be perfectly reconstructed at the output. The salient feature of these data-driven approaches is their flexibility. They can be easily generalized to practical scenarios beyond the relatively simple setup considered in prior works.

Our recent work \cite{gbaf} considered the practical packet-based transmission and proposed a new DL-based feedback code design, dubbed generalized block attention feedback (GBAF) code. Compared with existing DL-based code designs, GBAF code reduces the communication overhead, enables a wider range of code rates, and more importantly, provides a structured code design that achieves an ultra-low block error rate (BLER) even in the low-SNR regime. In this paper, we improve upon the GBAF code by endowing active feedback capability to it. The resulting feedback code, dubbed, \textit{block attention active feedback (BAAF) code}, outperforms GBAF by a large margin, achieving state-of-the-art BLER performance in the low-SNR regime.

\section{Preliminaries and Definitions}

\begin{figure*}[t]
\begin{center}
\includegraphics[scale =0.37]{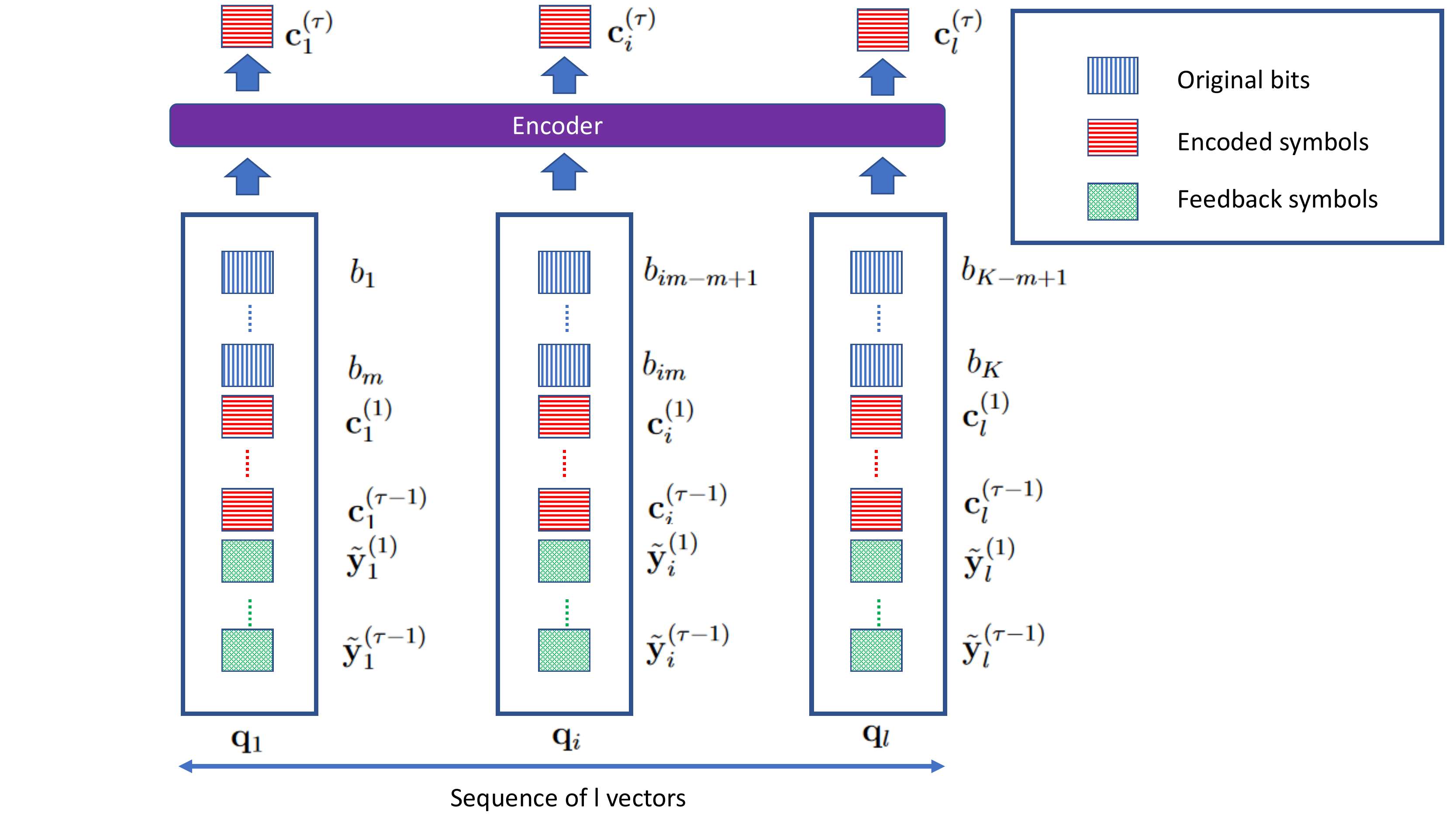}
\caption{Visualisation of sequence-to-sequence encoding with block structure at iteration $\tau$ for a block size of $m$.}
\label{seq2seq}
\end{center}
\end{figure*}


To reduce the feedback overhead in practical systems, we consider a \textit{block feedback} model, where the receiver actively codes and sends back a feedback signal after the transmission of a block of symbols in the forward direction.  
During communication block $\tau$ in the forward direction, the transmitter sends a block of $N_{\tau}$ symbols, denoted by $\mathbf{c}^{(\tau)}$, over the forward channel. Then, during communication block $\tau$ in the feedback direction, the receiver sends a block of $\tilde{N}_{\tau}$ symbols, denoted by $\tilde{\mathbf{c}}^{(\tau)}$, for $\tau=1,\ldots,T-1$. The communication is terminated when the receiver receives $\mathbf{c}^{(T)}$. 
Both the forward and feedback channels are modeled as additive white Gaussian noise (AWGN) channels. Accordingly, the received symbols over the forward and feedback channels, denoted by $\mathbf{y}^{(\tau)}$ and $\tilde{\mathbf{y}}^{(\tau)}$, respectively, are given by
\begin{equation}
\mathbf{y}^{(\tau)} = \mathbf{c}^{(\tau)} + \mathbf{n}^{(\tau)},
\end{equation}
and
\begin{equation}
\tilde{\mathbf{y}}^{(\tau)} = \tilde{\mathbf{c}}^{(\tau)} + \tilde{\mathbf{n}}^{(\tau)},
\end{equation}
where $\mathbf{n}^{(\tau)}\in\mathbb{R}^{N_{\tau}}$ and  $\tilde{\mathbf{n}}^{(\tau)}\in \mathbb{R}^{\tilde{N}_{\tau}}$ are vectors of independent zero-mean Gaussian random variables  with variances $\sigma^{2}_{ff} \mathbf{I}$ and $\sigma^{2}_{fb} \mathbf{I}$, respectively.

Assume that the transmitter wants to send $K$ bits, $\mathbf{b}=[b_{1},\ldots,b_{K}]\in \left\{0,1\right\}^{K}$, to the receiver in $N$ channel uses. We impose a rate constraint of $R$, that is we must have $N \leq K/R$, 
or, equivalently,
\begin{equation}
\sum_{\tau=1}^{T}N_{\tau}\leq N.
\end{equation}
Similarly, a constraint on the feedback direction can be introduced as,
\begin{equation}
\sum_{\tau=1}^{T-1}\tilde{N}_{\tau}\leq \tilde{N}.
\end{equation}

If we employ $T$ communication blocks, the direction of  communication changes $2T-1$ times, which we use to quantify the communication overhead of the described feedback mechanism. The more the number of blocks, the higher the  overhead introduced. For example, if $T=1$ there is only one forward transmission of $N$ symbols. On the other hand, if $N_{\tau}=1$, $\forall \tau$, then we have $N$ blocks and the transmitter waits for feedback after each symbol. The latter is the model considered in most prior papers, but this would introduce significant delays and overheads in practical systems.

In the general form of the described protocol, $N_{\tau}$ and  $\tilde{N}_{\tau}$ can be chosen as desired. For example, one may want to keep $\tilde{N}_{\tau}$ small to avoid potential delays due to feedback. Nevertheless, we consider $N_{\tau}=\tilde{N}_{\tau}$ in the scope of this paper to simplify the notation and the analysis. This choice is also desired in order to align forward and feedback symbols for the encoding mechanism.

The focus of our research, similar to the previous works, is to design a mechanism for generating symbols in forward and feedback directions for each  communication block $\tau$, $\tau=1,\ldots,T-1$. Next, we introduce the  knowledge vectors $\boldsymbol{q}^{(\tau)}$ and $\tilde{\boldsymbol{q}}^{(\tau)}$, which are used to represent all the available information to the transmitter and receiver, respectively, during  communication block $\tau$. The knowledge vector at the transmitter, $\boldsymbol{q}^{(\tau)}$, consists of the original bit stream, previously transmitted symbols, and the received feedback symbols, i.e.,
\begin{equation}
\boldsymbol{q}^{(\tau)} =[\mathbf{b},\mathbf{c}^{(1)},\ldots,\mathbf{c}^{(\tau)}, \tilde{\mathbf{y}}^{(1)},\ldots,\tilde{\mathbf{y}}^{(\tau-1)}].
\end{equation}
The knowledge vector at the receiver is given by
\begin{equation}
\tilde{\boldsymbol{q}}^{(\tau)} =[\tilde{\mathbf{c}}^{(1)},\ldots,\tilde{\mathbf{c}}^{(\tau-1)}, \mathbf{y}^{(1)},\ldots,\mathbf{y}^{(\tau)}].
\end{equation}
Now, let  $M^{({\tau})}$ and $\tilde{M}^{({\tau})}$ denote the encoding mechanisms for the forward and feedback directions, respectively, such that at each communication block $\tau$ we have the following mappings
\begin{equation}
M^{({\tau})}:  \boldsymbol{q}^{(\tau)} \xrightarrow{} \mathbf{c}^{(\tau)}\in \mathbb{R}^{N_{\tau}},
\end{equation}
and 
\begin{equation}
\tilde{M}^{({\tau})}:  \tilde{\boldsymbol{q}}^{(\tau)} \xrightarrow{} \tilde{\mathbf{c}}^{(\tau)}\in \mathbb{R}^{\tilde{N}_{\tau}}.
\end{equation}
\indent Once the transmission of all the symbols are completed, a decoding function $D$ at the receiver predicts the original bit stream, i.e.,
\begin{equation}
D: \tilde{\boldsymbol{q}}^{(T)}\xrightarrow{} \hat{\mathbf{b}}\in\left\{0,1\right\}^{K}.
\end{equation}

\subsubsection{Power constraint}
The transmitter and receiver must satisfy average power constraints that can be formally described as:
\begin{align}
\mathbb{E}\left[\frac{1}{N} \sum^{T}_{\tau=1}\langle \mathbf{c}^{(\tau)} , \mathbf{c}^{(\tau)} \rangle\right]\leq 1,
\end{align}
and
\begin{align}
\mathbb{E}\left[\frac{1}{\tilde{N}} \sum^{T-1}_{\tau=1}\langle \tilde{\mathbf{c}}^{(\tau)} , \tilde{\mathbf{c}}^{(\tau)} \rangle\right]\leq 1.
\end{align}
We define the SNR of the forward channel as $SNR_{ff}=1/\sigma_{ff}^{2}$, and the feedback channel as $SNR_{fb}=1/\sigma_{fb}^{2}$.

\subsubsection{Active vs. Passive Feedback}
The model described above is called an {\ active feedback} channel as the encoder at the receiver, $\tilde{M}^{({\tau})}$, actively processes its knowledge vector $\tilde{\boldsymbol{q}}^{(\tau)}$ to generate the vector of symbols, $\tilde{\mathbf{c}}^{(\tau)}$, transmitted over the feedback channel. On the other hand, in the case of {\em passive feedback}, the encoder $\tilde{M}^{({\tau})}$ acts simply as a relaying mechanism, where
\begin{equation}\label{relay}
\tilde{M}^{({\tau})}:  \tilde{\boldsymbol{q}}^{(\tau)} \xrightarrow{relay} \tilde{\mathbf{c}}^{(\tau)} = \alpha\mathbf{y}^{(\tau)}=\alpha\mathbf{c}^{(\tau)}+\alpha\mathbf{n}^{(\tau)},
\end{equation}
where $\alpha$ is a scalar that scales the received vector $\mathbf{y}^{(\tau)}$ to satisfy the average power constraint. Hence, in the case of passive feedback, we always have $\tilde{N}_{\tau}=N_{\tau}$.

\subsubsection{Systematic Feedback}
We refer to a feedback mechanism as {\em systematic feedback}, if at $\tau=1$, the encoder at the transmitter maps the original bit stream to its BPSK modulated version, i.e.,
$N_{1}=K$, and
\begin{equation}\label{encr}
M^{({1})}:  \boldsymbol{q}^{(1)}=\mathbf{b} \xrightarrow{BPSK} \mathbf{c}^{(1)} = \bar{\mathbf{b}}=2*\mathbf{b}-1,
\end{equation}
and the encoder at the receiver simply relays the received noisy symbols, i.e.,
$\tilde{N}_{1}=K$, and
\begin{equation}\label{enct}
\tilde{M}^{({1})}:  \tilde{\boldsymbol{q}}^{(1)}=\mathbf{y}^{(1)} \xrightarrow{relay} \tilde{\mathbf{c}}^{(1)} = \bar{\mathbf{b}}+\mathbf{n}^{(1)}.
\end{equation}

Note that this notion can be extended to modulation schemes with larger constellations. Although the proposed GBAF code does not employ systematic encoding, we introduced it as it has been the standard choice in previous works studying DL-aided code design for feedback channels \cite{deepcode, defc, drf}.

\begin{figure*}[t]
\begin{center}
\includegraphics[scale =0.35]{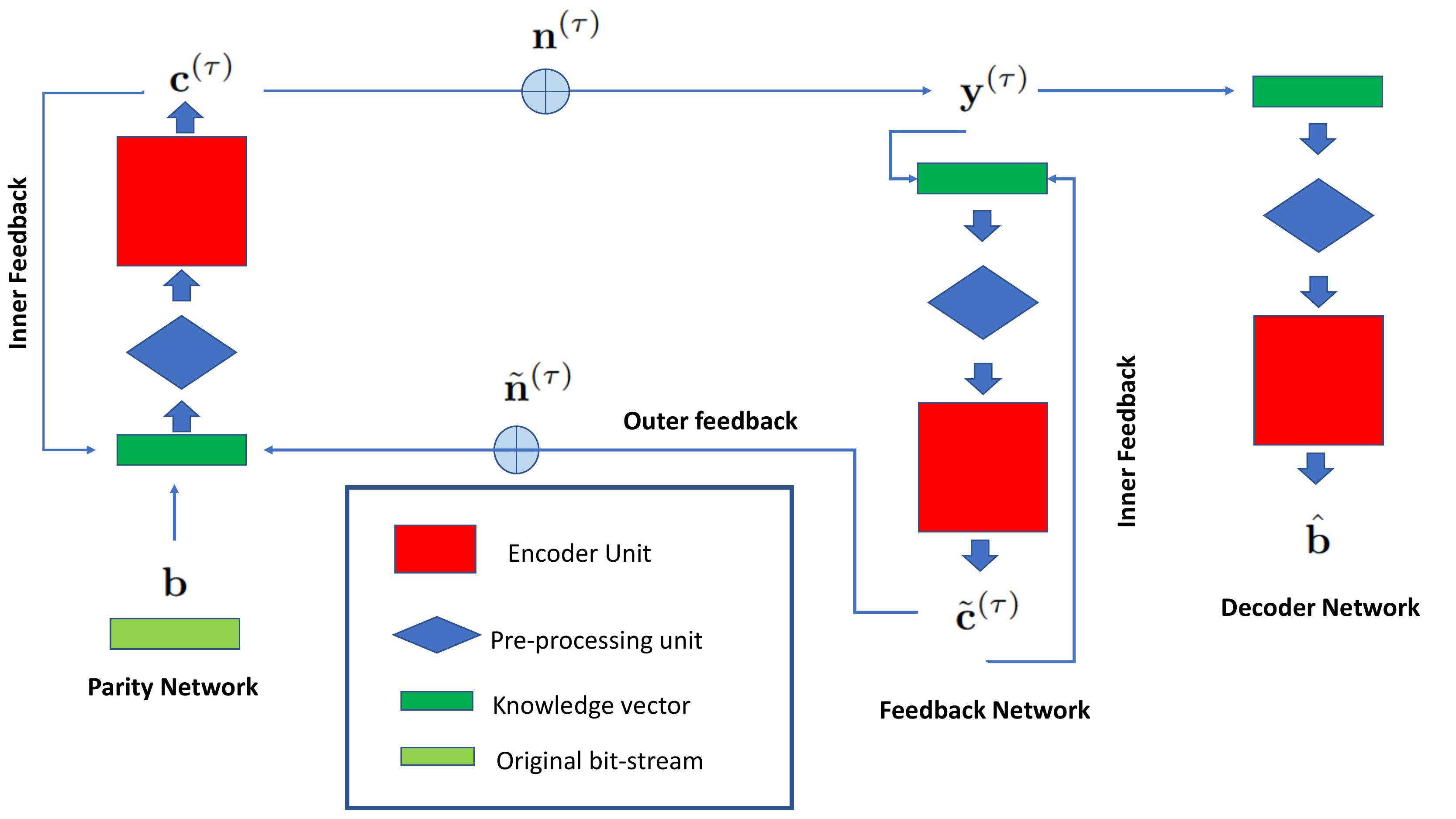}
\caption{Illustration of the overall common code architecture used by both the GBAF and BAAF codes. The green, blue, and red blocks denote the knowledge vector, pre-processing unit and encoder unit, respectively. The dashed lines and shapes indicate the units and connections that are optional.}
\label{sysmodel}
\end{center}
\end{figure*}
\section{Block Attention Active Feedback (BAAF) Codes}

\subsection{Block Attention Feedback Codes}
GBAF codes that were introduced in \cite{gbaf} aim to generate parity symbols to communicate by using a transformer-based architecture \cite{Wang:transformer:comms} within a sequence-to-sequence encoding framework. Before the technical discussions, we want to explain certain concepts to better highlight the connection between the {\em communication with feedback problem} and the transformer architecture used for the code design. One of the critical concepts is the {\em sequence}, which we use to refer to an input structure that consists of a certain number of ordered elements, for instance, in the case of natural language processing (NLP) tasks, a sentence is a sequence of words, and each word is an element of this sequence. Accordingly, in the context of communication, we treat a bit-stream as a sequence of bits; however, instead of treating each bit as a distinct element, we divide the $K$ information bits into $l$ blocks of $m$ bits each, $K=l \cdot m$, to have a sequence of $l$ vectors, i.e., $\left\{\bm{b}_{1},\ldots,\bm{b}_{l}\right\}$.

Accordingly, to generate coded symbols, knowledge vectors are also organised into a sequence of length $l$, $\mathcal{Q}=b \left\{\bm{q}_{1}, \ldots, \bm{q}_{l} \right\}$, where $\bm{q}_{i}$ represents the knowledge sub-vector corresponding to the subset of bits $\bm{b}_{i}$, $i=1, \ldots, l$. Initially we have $\bm{q}^{(1)}_{i}=\bm{b}_{i}$. These knowledge vectors are fed into the DNN architecture, output of which is also a sequence of $l$ vectors containing coded symbols, i.e., $\left\{\bm{c}_{1},\ldots,\bm{c}_{l}\right\}$. 

\indent Once the vector of coded symbols, $\bm{c}^{(\tau)}=\left[  \bm{c}^{(\tau)}_{1},\ldots,\bm{c}^{(\tau)}_{l} \right]$, is transmitted over the forward channel in block $\tau$, and the corresponding vector of feedback symbols, $\tilde{\bm{c}}^{(\tau)}=\left[ \tilde{\bm{c}}^{(\tau)}_{1},\ldots,\tilde{\bm{c}}^{(\tau)}_{l} \right]$ is received, the elements of the knowledge vector are updated as follows:\footnote{In the implementation, since the DNN architecture requires fixed vector size we use zero padding for later coded symbols and feedback symbols when needed.} 
\begin{equation}
\bm{q}^{(\tau+1)}_{i}=[\mathbf{b}_{i},\mathbf{c}^{(1)}_{i},\ldots,\mathbf{c}^{(\tau)}_{i}, \tilde{\mathbf{y}}^{(1)}_{l},\ldots,\tilde{\mathbf{y}}^{(\tau)}_{i}].
\end{equation}
 We illustrate the encoding process with the explained sequence-to-sequence encoding framework with the block structure utilized in GBAF codes in Fig. \ref{seq2seq}.

\subsection{Active Feedback Architecture}
For the active feedback code design, we follow the same architectural principles introduced in \cite{gbaf}. The overall architecture consists of three network blocks; namely, {\em parity network}, {\em feedback network}, and 
{\em decoder network}, as illustrated in Fig \ref{sysmodel}. The parity network and the feedback network operate sequentially to generate a sequence of parity symbols $\bm{c}^{(\tau)}=\left[  \bm{c}^{(\tau)}_{1},\ldots,\bm{c}^{(\tau)}_{l} \right]$ and the corresponding feedback symbols $\tilde{\bm{c}}^{(\tau)}=\left[ \tilde{\bm{c}}^{(\tau)}_{1},\ldots,\tilde{\bm{c}}^{(\tau)}_{l} \right]$, respectively. The decoder network operates at the end of $T$ communication blocks at the receiver, and provides a prediction $\hat{\mathbf{b}}$ for the original bit stream. 

For the parity and feedback networks, we utilize two feedback mechanisms; namely {\em inner feedback} and  {\em outer feedback}, as illustrated in Fig \ref{sysmodel}. The outer feedback enables the interaction between two networks through coded parity and feedback symbols transmitted over the channel. On the other hand, the inner feedback is used to allow each network to track its previously generated coded symbols. Different from parity network and feedback network, decoder network does not require a feedback mechanism since operates only once at the very end.

Each network consists of two main components, an encoder unit and a pre-processing unit, denoted by $H_{encoder}(\cdot)$ and $S(\cdot)$, respectively. The objective of the pre-processing unit is to convert the knowledge vector $\boldsymbol{q}$ into a sequence of knowledge vectors, i.e., 
$S(\boldsymbol{q})=\left\{\bm{q}_{1},\ldots,\bm{q}_{l}\right\}$. The encoder unit $H_{encoder}(\cdot)$ is responsible for mapping the sequence of knowledge vectors to either coded symbols (for parity and feedback networks) or logits (for decoder) network, i.e.,
\begin{equation}
H_{\mathrm{encoder}}: \mathcal{Q}=\left\{\bm{q}_{1},\ldots,\bm{q}_{l} \right\}\xrightarrow{Encoding} \mathcal{U}=\left\{\bm{u}_{1},\ldots,\bm{u}_{l} \right\},
\end{equation}
such that $\bm{q}_{i}\in\mathbb{R}^{d_{in}}$, $\bm{u}_{i}\in\mathbb{R}^{d_{out}}$.\\
\indent $H_{\mathrm{encoder}}$ unit consists of sequential combination of three main modules, in the following order: feature extractor $H_{\mathrm{extract}}$, sequence-to-sequence encoder $H_{s2s}$, and output mapping $H_{map}$.

\subsection{Modules Used in the Architecture}

We next describe some of the modules we use in the BAAF code architecture. 

\subsubsection{Feature extractor}
The feature extractor $H_{\mathrm{extract}}$ has two main objectives. First, it is used to map $d_{in}$-dimensional knowledge vectors to a certain vector representation of size $d_{model}$ that $H_{s2s}$ accepts as an input. Second, it helps to keep the input that is fed to $H_{s2s}$ within a certain range. Since the raw data contains noisy values, particularly in the low SNR regime, a simple linear mapping may overemphasize noise terms. Hence, for $H_{\mathrm{extract}}$, we use the MLP architecture proposed in \cite{gbaf}, where we replace the GeLU activation with ReLU activation. The structure of $H_{\mathrm{extract}}$ is identical for all three networks.

\subsubsection{Sequence-to-sequence encoder}
The sequence-to-sequence encoder $H_{s2s}$ is the core of the encoding mechanism. It processes a sequence of $l$ $d_{model}$-dimensional vectors, and maps them to a sequence of $l$ $d_{model}$-dimensional vectors, such that each vector in the latter sequence contains certain information about the other vectors in the sequence. For sequence-to-sequence encoding, we employ the transformer architecture. We remark here that, although the sequence-to-sequence encoding can also be performed by other neural network architectures, such as LSTM \cite{lstm} and GRU \cite{gru}, recent works have shown that self-attention-based transformer architecture outperforms these recurrent-based alternatives.

Our $H_{s2s}$ module is a stack of $N$ identical {\em transformer encoder layers}\footnote{We follow the standard implementation used in the Pytorch library:\\ \url{https://pytorch.org/docs/stable/_modules/torch/nn/modules/transformer.html#TransformerEncoderLayer}}, where each layer consists of three main submodules: feed-forward module, multi-head attention module, and layer normalization module. Since we have employed the standard transformer encoder layer architecture here, we do not provide further details, and instead refer the reader to \cite{gbaf} and references therein. For the implementation, following \cite{gbaf}, we set $d_{model}=32$, $N_{parity}=2$, $N_{feedback}=2$ and $N_{decoder}=3$.

\begin{algorithm}[t]
\caption{Iterative parity symbol encoding (IPSE)}\label{alg:ipse}
\begin{algorithmic}[1]
\For{$\tau=1,\ldots,T$}
\State{{\color{red}\textbf{Transmitter}}:}
\State{\textbf{Update knowledge vector}:}
\State{$\boldsymbol{q}^{(\tau)}=[\bm{b},\bm{c}^{(1)},\ldots,\bm{c}^{(\tau-1)}, \tilde{\bm{y}}^{(1)},\ldots,\tilde{\bm{y}}^{(\tau-1)}]$}
\State{\textbf{Pre-process knowledge vector}:}
\State{$\left\{\bm{q}^{(\tau)}_{i},\ldots,\bm{q}^{(\tau)}_{l}\right\}=S_{parity}(\boldsymbol{q}^{(\tau)})$}
    \State{\textbf{Feature extraction:}}
    \For{$i\in[l]$}~~$\bm{f}^{(\tau)}_{i} = H^{parity}_{extract}(\bm{q}^{(\tau)}_{i})$
    \EndFor
    \State{\textbf{Attention-based neural-encoding}:}
    \State{$ \mathcal{V}^{(\tau)}= H^{parity}_{s2s}(\mathcal{F}^{(\tau)})$}
    \State{\textbf{Symbol mapping}:}
    \For{$i\in[l]$}~~$c^{(\tau)}_{i}= H^{parity}_{map}(\bm{v}^{(\tau)}_{i})$
    \EndFor
\State{{\color{blue}\textbf{Receiver}}:}
\State{\textbf{Update knowledge vector}:}
\State{$\tilde{\boldsymbol{q}}^{(\tau)}=[\tilde{\bm{c}}^{(1)},\ldots,\tilde{\bm{c}}^{(\tau-1)}, \bm{y}^{(1)},\ldots,\bm{y}^{(\tau)}]$}
\State{\textbf{Pre-process knowledge vector}:}
\State{$\left\{\tilde{\bm{q}}^{(\tau)}_{1},\ldots,\tilde{\bm{q}}^{(\tau)}_{l}\right\}=S_{feedback}(\tilde{\boldsymbol{q}}^{(\tau)})$}
    \State{\textbf{Feature extraction:}}
    \For{$i\in[l]$}
    \State{$\tilde{\bm{f}}^{(\tau)}_{i} = H^{feedback}_{extract}(\tilde{\bm{q}}^{(\tau)}_{i})$}
    \EndFor
    \State{\textbf{Attention-based neural-encoding}}:
    \State{$ \tilde{\mathcal{V}}^{(\tau)}= H^{feedback}_{s2s}(\tilde{\mathcal{F}}^{(\tau)})$}
    \State{\textbf{Symbol mapping}:}
    \For{$i\in[l]$}
    \State{$\tilde{c}^{(\tau)}_{i}= H^{feedback}_{map}(\tilde{\bm{v}}^{(\tau)}_{i})$}
    \EndFor
\EndFor
\end{algorithmic}
\end{algorithm}

\subsubsection{Output mapping}
Let $\mathcal{V}=\left\{\bm{v}_{1},\ldots,\bm{v}_{l} \right\}$ be the output sequence of $H_{s2s}$ and the final latent representations. The role of the output mapping $H_{map}$ is  to map the final latent representation, i.e.,
$H_{\mathrm{map}}(\bm{v}_{i})=\bm{u}_{i}$, for all $i\in[l]$, where $\bm{u}_{i}\in\mathbb{R}^{d_{out}}$. For all three networks, we use a common structure of a fully connected layer for $H_{map}$; however, for each network we consider different $d_{out}$ values. For parity and feedback networks $\bm{u}_{i}=[c^{\tau}_{i}]$ and $\bm{u}_{i}=[\tilde{c}^{\tau}_{i}]$, respectively, thus for these networks we have $d_{out}=1$. On the other hand, the decoder network performs classification for each block of m-bits; that is, it maps the latent representation to one of the $2^{m}$ possible $m$-length bit streams. Hence, for $H^{parity}_{map}$, we have $d_{out}=2^{m}$. We also note that in the case of $H^{decoder}_{map}$, the fully connected layer is followed by a softmax layer to obtain $2^{m}$ dimensional logit vector. Finally, we note here that, due to the average power constraint, an extra layer for power normalization is required following the $H^{parity}_{map}$ and $H^{feedback}_{map}$, for which we follow the same procedure used previously in \cite{deepcode, AttentionCode, gbaf}.

\subsection{Algorithmic Flow and Training Procedure}
In this subsection, we provide the algorithmic flow of the proposed BAAF code design to highlight the important steps and described the overall end-to-end encoding and decoding mechanisms. The overall encoding-decoding process consists of two algorithmic flows: {\em iterative parity symbol encoding (IPSE)} and {\em joint parity symbol decoding (JPSD)}.

IPSE, summarized in Algorithm \ref{alg:ipse}, defines the interactive communication between the receiver and transmitter. The algorithm consists of 5 main steps for each of the transmitter and the receiver: i) updating of the knowledge vector (lines 4 and 15), ii) pre-processing of the knowledge vectors (lines 6 and 17), iii) feature extraction (lines 8 and 19), iv) attention-based neural encoding (lines 10 and 21), and v) symbol mapping (lines 12 and 23).

Once the interactions between the transmitter and the receiver are completed, at the end of the $T$th communication block, the receiver executes JPSD, summarized in Algorithm \ref{alg:jpsd}. The steps of JPSD are almost identical to those of IPSE, except the last one, where the latent representation is mapped to a logit vector to predict the correct bit-block representation among $2^{m}$ candidates. Hence, from the decoder's perspective, the objective can be considered as a multi-label classification task with $2^{m}$ labels. For training, we first randomly generate a sequence of bits $\bm{b}\in\left\{0,1\right\}^{K}$, then divide it into $l$ blocks, each of size $m$ bits, and assign  the corresponding label for each block to obtain the training data $\mathcal{B}= \left\{(\bm{b}_{1},y_{1}), \ldots, (\bm{b}_{l},y_{l})\right\}$. Then, the generated data of bit-blocks is fed into the IPSE algorithm, and its results are further processed by JPSD to output a sequence of $2^{m}$-dimensional vector of length $l$,  $\left\{\bm{w}_{1}, \ldots,\bm{w}_{l}\right\}$ (line 9, Algorithm \ref{alg:jpsd}). We consider cross-entropy loss for the corresponding multi-label classification problem, i.e., 
\begin{equation}\label{loss}
L(\bm{W},Y) = \sum^{l}_{i=1}\sum^{2^{m}-1}_{c=0}-\log\frac{\exp{(\bm{W}_{[i,c]})}}{\sum^{2^{m}-1}_{c=0}\exp{(\bm{W}_{[i,c]})}}\cdot\mathbbm{1}_{y_{i}\neq c}
\end{equation}
where $Y=\left\{y_{1},\ldots,y_{l}\right\}$ is the list of labels of blocks in the generated sequence, and $\bm{W}$ is the matrix form of the sequence $\left\{\bm{w}_{1},\ldots,\bm{w}_{l}\right\}$. For further technical details please refer to \cite{gbaf}.

\begin{algorithm}[t!]
\caption{Joint parity symbol decoding (JPSD)}\label{alg:jpsd}
\begin{algorithmic}[1]
\State{\textbf{Update Knowledge vector}:}
\State{$\hat{\boldsymbol{q}} =[\tilde{\bm{c}}^{(1)},\ldots,\tilde{\bm{c}}^{(T-1)}, \bm{y}^{(1)},\ldots,\bm{y}^{(T)}]$}
\State{\textbf{Pre-process knowledge vector for decoder network}:}
\State{$S_{\mathrm{decoder}}(\hat{\boldsymbol{q}})=\left\{\hat{\bm{q}}_{1},\ldots,\hat{\bm{q}}_{l}\right\}$,~~ $\hat{\bm{q}}_{i}=[\tilde{\bm{y}}^{(1)}_{i},\ldots,\tilde{\bm{y}}^{(T)}_{i}])$}
\State{\textbf{Feature extraction:}}
\For{$i\in[l]$}~~$\hat{\bm{f}}_{i} = H^{decoder}_{extract}(\hat{\bm{q}}_{i})$
\EndFor
\State{\textbf{Attention-based neural-encoding}:$ \hat{\mathcal{V}}= H^{decoder}_{s2s}(\hat{\mathcal{F}})$}
\State{\textbf{Mapping}:}
\For{$i\in[l]$}
$\bm{w}_{i}= H^{decoder}_{map}(\hat{\bm{v}}_{i})$
\EndFor
\State{\textbf{Block-wise classification}:} 
\For{$i\in[l]$} $p_{i}= \max_{j}(\bm{w}_{i})_{[j]}$
\EndFor
\State{\textbf{Block index to bitstream conversion}:}
\For{$i\in[l]$} $\hat{\bm{b}}=[\hat{\bm{b}},\bm{A}_{[p_{i},:]}]$
\EndFor
\end{algorithmic}
\end{algorithm}

\section{Numerical Results}\label{s:experiments}

In all the experiments, we consider a bit stream of length $K=51$. For block attention feedback code designs we fixed the block size to $m=3$, such that the number of blocks is $l=17$. We consider $SNR_{ff}\in[-1,2]$~dB, and the transmission of $T=9$ parity bits for each block in total, which corresponds to a transmission rate of $R=3/9 = 1/3$. For the feedback channel, we also consider additive Gaussian noise with $SNR_{fb}=20$dB.

For training, we utilized the AdamW optimizer, which is a variation of the Adam optimizer with decoupled weight decay regularization \cite{adamw}. We consider a batch size of $B=8192$, an initial learning rate of $0.001$, and a weight decay parameter $0.01$. In addition, we apply gradient clipping with a threshold of $0.5$. We train the network for $140K$ batches using cross-entropy loss and apply polynomial decay to the learning rate. We also employ curriculum learning; that is, in the first 20000 batches we gradually decrease the feed-forward channel SNR starting from 3dB while fixing the feedback channel SNR at 100dB. Then, during the second 20000 batches, we gradually decrease the feedback SNR from 100dB to 20 dB.

We remark that during training power normalization is performed over batches to satisfy the average power constraints; hence, at the test time we first obtain the statistics over a batch and then fix the statistics (mean and variance) for power normalization.

\begin{figure}[t]
\begin{center}
\includegraphics[scale =0.6]{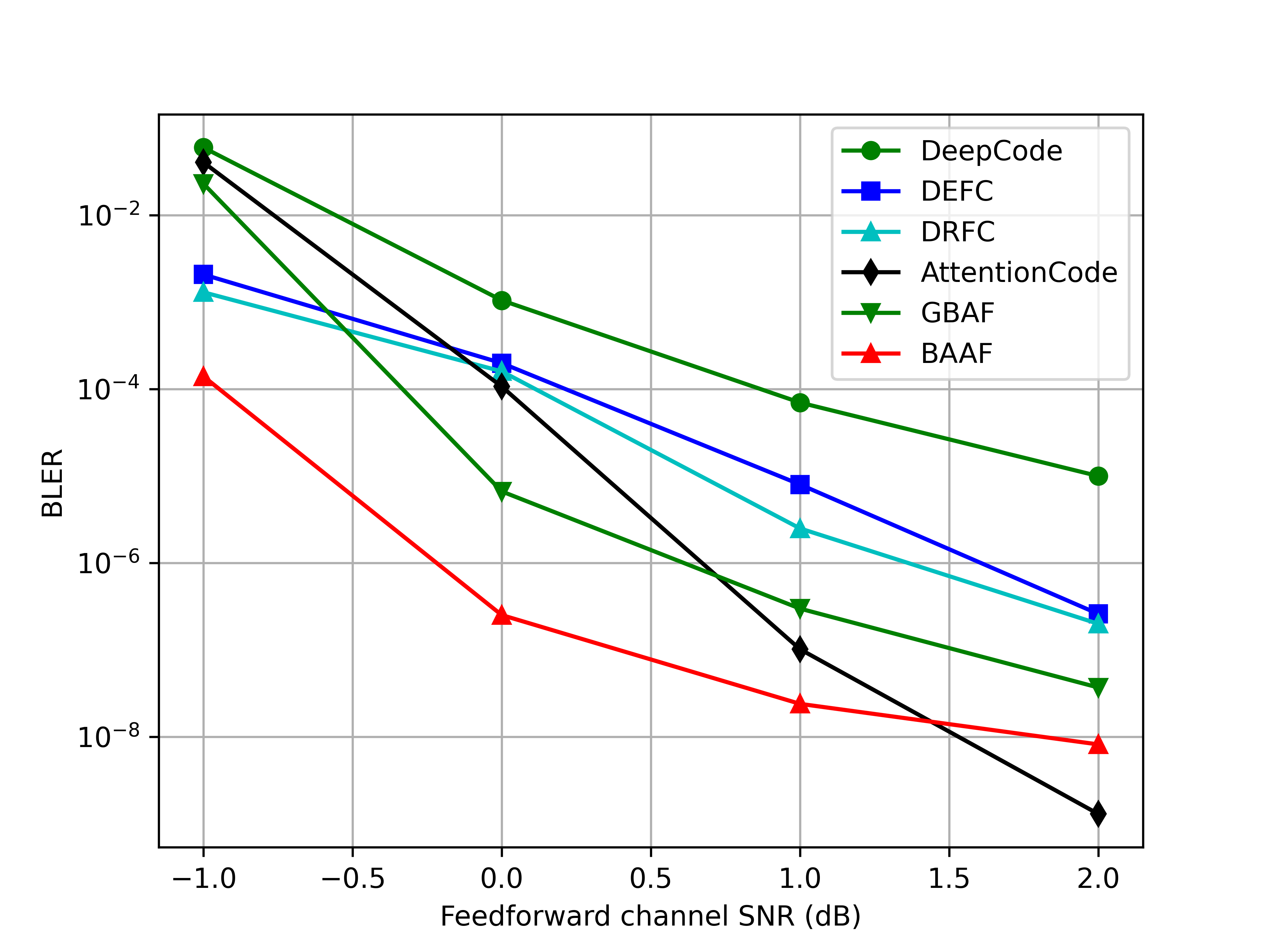}
\caption{BLER performance of BAAF codes benchmarked against state-of-the-art alternatives, which also benefit from DNN-based architectures and data-driven optimization.}
\label{comp}
\end{center}
\end{figure}

For comparison we consider the BLER as the performance measure\footnote{For the reliability of the results, we run the codes until at least 100 errors are observed.} and compare the proposed BAAF design with the existing DNN-based feedback code alternatives that focus on passive feedback: GBAF code \cite{gbaf}, DeepCode \cite{deepcode}, DEFC \cite{defc}, DRFC \cite{drf}, and AttentionCode \cite{AttentionCode}. The results are presented in Fig. \ref{comp}. The results show that active feedback further improves the performance of the block-attention feedback design significantly, especially in the low SNR regime. The results also indicate that at forward channel SNR values of $-1,0,1$dB, the BAAF code outperforms all other previous designs. We highlight that, while the AttentionCode outperforms BAAF in the high SNR regime, it is based on more frequent interactions between the receiver and transmitter; that is, it has a block size of $1$, which corresponds to a significantly higher feedback overhead.

\section{Conclusion}\label{s:Conclusion}
In this work, we have introduced BAAF codes for reliable communication over active feedback channels. In BAAF codes, both the parity and feedback symbols are sequentially generated by using transformer-based neural network architectures. We have shown that, thanks to the active feedback mechanism at the receiver, BAAF codes can further reduce the BLER performance compared to the state-of-the-art passive feedback code designs, especially in the low SNR regime.

\bibliographystyle{IEEEtran}
\bibliography{IEEEabrv,ref.bib}

\begin{thebibliography}{10}
\providecommand{\url}[1]{#1}
\csname url@samestyle\endcsname
\providecommand{\newblock}{\relax}
\providecommand{\bibinfo}[2]{#2}
\providecommand{\BIBentrySTDinterwordspacing}{\spaceskip=0pt\relax}
\providecommand{\BIBentryALTinterwordstretchfactor}{4}
\providecommand{\BIBentryALTinterwordspacing}{\spaceskip=\fontdimen2\font plus
\BIBentryALTinterwordstretchfactor\fontdimen3\font minus
  \fontdimen4\font\relax}
\providecommand{\BIBforeignlanguage}[2]{{%
\expandafter\ifx\csname l@#1\endcsname\relax
\typeout{** WARNING: IEEEtran.bst: No hyphenation pattern has been}%
\typeout{** loaded for the language `#1'. Using the pattern for}%
\typeout{** the default language instead.}%
\else
\language=\csname l@#1\endcsname
\fi
#2}}
\providecommand{\BIBdecl}{\relax}
\BIBdecl

\bibitem{gallager1962low}
R.~Gallager, ``Low-density parity-check codes,'' \emph{IRE Transactions on
  information theory}, vol.~8, no.~1, pp. 21--28, 1962.

\bibitem{berrou1993near}
C.~Berrou, A.~Glavieux, and P.~Thitimajshima, ``Near {S}hannon limit
  error-correcting coding and decoding: Turbo-codes,'' in \emph{IEEE ICC},
  1993.

\bibitem{arikan2009channel}
E.~Arikan, ``Channel polarization: A method for constructing capacity-achieving
  codes for symmetric binary-input memoryless channels,'' \emph{IEEE Trans. on
  Inform. Theory}, vol.~55, no.~7, pp. 3051--3073, 2009.

\bibitem{AlphaSeq}
Y.~Shao, S.~C. Liew, and T.~Wang, ``Alphaseq: sequence discovery with deep
  reinforcement learning,'' \emph{IEEE Trans. Neural Netw. Lear. Syst.},
  vol.~31, no.~9, pp. 3319--3333, 2019.

\bibitem{Shannon}
C.~Shannon, ``The zero error capacity of a noisy channel,'' \emph{IRE
  Trans.Inf. Theory}, vol.~2, no.~3, pp. 8--19, 1956.

\bibitem{SK1}
J.~Schalkwijk and T.~Kailath, ``A coding scheme for additive noise channels
  with feedback i: No bandwidth constraint,'' \emph{IEEE Trans. Inf. Theory},
  vol.~12, no.~2, pp. 172--182, 1966.

\bibitem{SK2}
J.~Schalkwijk, ``A coding scheme for additive noise channels with feedback ii:
  Band-limited signals,'' \emph{IEEE Trans. Inf Theory}, vol.~12, no.~2, pp.
  183--189, 1966.

\bibitem{Kim2007ISIT}
Y.-H. Kim, A.~Lapidoth, and T.~Weissman, ``The {G}aussian channel with noisy
  feedback,'' in \emph{IEEE ISIT}, 2007, pp. 1416--1420.

\bibitem{ModuloSK}
A.~Ben-Yishai and O.~Shayevitz, ``Interactive schemes for the {AWGN} channel
  with noisy feedback,'' \emph{IEEE Trans. Inf. Theory}, 2017.

\bibitem{Gallager2}
B.~Nakibo{\u{g}}lu and R.~G. Gallager, ``Error exponents for variable-length
  block codes with feedback and cost constraints,'' \emph{IEEE Transactions on
  Information Theory}, vol.~54, no.~3, pp. 945--963, 2008.

\bibitem{Sahai}
A.~Sahai, ``Why do block length and delay behave differently if feedback is
  present?'' \emph{IEEE Transactions on Information Theory}, vol.~54, no.~5,
  pp. 1860--1886, 2008.

\bibitem{ozarow1984achievable}
L.~Ozarow and S.~Leung-Yan-Cheong, ``An achievable region and outer bound for
  the {G}aussian broadcast channel with feedback,'' \emph{IEEE Trans. Info.
  Theory}, vol.~30, no.~4, pp. 667--671, 1984.

\bibitem{ozarow1984capacity}
L.~Ozarow, ``The capacity of the white {G}aussian multiple access channel with
  feedback,'' \emph{IEEE Transactions on Information Theory}, vol.~30, no.~4,
  pp. 623--629, 1984.

\bibitem{bross2008relay}
S.~Bross and M.~Wigger, ``On the relay channel with receiver--transmitter
  feedback,'' \emph{IEEE Trans. Info. Theory}, vol.~55, no.~1, pp. 275--291,
  2008.

\bibitem{deepcode}
H.~Kim, Y.~Jiang, S.~Kannan, S.~Oh, and P.~Viswanath, ``Deepcode: Feedback
  codes via deep learning,'' \emph{IEEE Journal on Selected Areas in
  Information Theory}, vol.~1, no.~1, pp. 194--206, 2020.

\bibitem{defc}
A.~Safavi, A.~Perotti, B.~Popovic, M.~Mashhadi, and D.~Gunduz, ``Deep extended
  feedback codes,'' \emph{ITU Journal on Future and Evolving Technologies},
  vol.~2, no.~6, pp. 33--41, 2021.

\bibitem{drf}
M.~B. Mashhadi, D.~G{\"{u}}nd{\"{u}}z, A.~Perotti, and B.~M. Popovic, ``{DRF}
  codes: Deep {SNR}-robust feedback codes,'' \emph{arXiv 2112.11789}, 2021.

\bibitem{AttentionCode}
Y.~Shao, E.~Ozfatura, A.~Perotti, B.~Popovic, and D.~Gunduz, ``Attentioncode:
  Ultra-reliable feedback codes for short-packet communications,''
  \emph{arXiv:2205.14955}, 2022.

\bibitem{gbaf}
E.~Ozfatura, Y.~Shao, A.~Perotti, B.~Popovic, and D.~Gunduz, ``All you need is
  feedback: Communication with block attention feedback codes,''
  \emph{arXiv:2206.09457}, 2022.

\bibitem{Wang:transformer:comms}
Y.~Wang, Z.~Gao, D.~Zheng, S.~Chen, D.~Gunduz, and V.~Poor,
  ``Transformer-empowered {6G} intelligent networks: From massive {MIMO}
  processing to semantic communication,'' \emph{IEEE Wireless Comms. Mag.},
  2022.

\bibitem{lstm}
H.~Sak, A.~W. Senior, and F.~Beaufays, ``Long short-term memory based recurrent
  neural network architectures for large vocabulary speech recognition,''
  \emph{CoRR}, vol. abs/1402.1128, 2014.

\bibitem{gru}
K.~Cho, B.~van Merrienboer, D.~Bahdanau, and Y.~Bengio, ``On the properties of
  neural machine translation: Encoder-decoder approaches,'' \emph{CoRR}, vol.
  abs/1409.1259, 2014.

\bibitem{adamw}
I.~Loshchilov and F.~Hutter, ``Decoupled weight decay regularization,'' in
  \emph{International Conference on Learning Representations}, 2019.

\end{thebibliography}

\end{document}